\pdfoutput=1
\documentclass[12pt,notitlepage,twocolumn,twoside]{article}
\usepackage{times}

\usepackage{times}
\usepackage{mathptmx}

\newcommand{\xx}[1]{\section {#1}}
\newcommand{\xxx}[1]{\subsection {#1}}
\newcommand{\xxxx}[1]{\subsubsection {#1}}

\providecommand{\spacing}[1]{\renewcommand{\baselinestretch}{#1}\small\normalsize}
\providecommand{\doublespacing}{\spacing{1.5}}
\providecommand{\normalspacing}{\spacing{1.0}}

\newcommand{\eql}[2]{\begin{equation} \label{eq:#1} #2 \end{equation}}
\newcommand{\eqr}[1]{Eq.~(\ref{eq:#1})}

\newcommand{\secl}[1]{\label{sec:#1}}
\newcommand{\secr}[1]{\S\,\ref{sec:#1}}

\newcommand{\figr}[1]{Fig.~\protect{\ref{fig:#1}}}

\newcommand{\tabr}[1]{Table~\protect{\ref{tab:#1}}}

\newcommand{\tbox}[1]{\begin{tabular}{c} #1 \end{tabular}}
\newcommand{\ylab}[1]{\tbox{\rotatebox{90}{#1}}}

\usepackage[normalem]{ulem}
\newcommand{\degrees}[1]{\mbox{$ {#1}^\circ $}}
\newcommand{\degreesC}[2][~]{\mks[\degrees{{#1}}]{{#2}}{C}}
\newcommand{\dvar}[1]{{#1}d-$\!$VAR}
\newcommand{\eg}{{\em e.g.}}

\newcommand{\etal}{{\em et~al.}}

\newcommand{\ie}{{\em i.e.}}

\newcommand{\J}[1]{\mbox{$ J_{#1} $}}

\newcommand{\hr}[2][~]{\mks[{#1}]{{#2}}{h}}
\newcommand{\hrs}[2][~]{\mks[{#1}]{{#2}}{h}}
\newcommand{\km}[2][~]{\mks[{#1}]{{#2}}{km}}

\newcommand{\mks}[3][~]{\mbox{$\mathrm{{#2}\mbox{{#1}}{#3}}$}}

\newcommand{\ms}[2][~]{\mks[{#1}]{{#2}}{m \; s^{-1}}}

\newcommand{\Vector}[1]{\mbox{$\mathbf{{#1}}$}}
\newcommand{\wrt}{with respect to}

\newcommand{\tophline}{\hline\noalign{\vspace{1mm}}}
\newcommand{\middlehline}{\noalign{\vspace{1mm}}\hline\noalign{\vspace{1mm}}}
\newcommand{\bottomhline}{\noalign{\vspace{1mm}}\hline}

\usepackage{ametsoc}

\renewcommand{\secr}[2][~]{Sect.{{#1}}\ref{sec:#2}}

\usepackage[final]{graphicx}
\graphicspath{ {/users/rhoffman/NWP/fig/misc/} }
\DeclareGraphicsExtensions{.pdf,.png,.jpeg}

\newcommand{\tfbox}[1]{\fbox{\begin{tabular}{c} #1 \end{tabular}}}
\newcommand{\tpbox}[1]{\parbox[c]{5.5in}{\vspace*{5pt}{#1}\vspace*{5pt}}}

\newcommand{\fitem}[3]{\begin{figure}[htb]\vspace*{2mm}
 \centerline{{\sffamily\upshape\bfseries {#3}}} 
 \caption{{#2}\label{fig:#1}}
\end{figure}}

\newcommand{\fitemstar}[3]{\begin{figure*}[tbh]\vspace*{2mm}
 \centerline{{\sffamily\upshape\bfseries {#3}}} 
 \caption{{#2}\label{fig:#1}}
\end{figure*}}

\begin{document}

\onecolumn

\title{\vspace*{-1in}Potential of \dvar4 for exigent forecasting of severe weather$^{1}$}

\author{Ross N. Hoffman$^{2,3}$, John
M. Henderson$^{2}$, and Thomas Nehrkorn$^{2}$}

\footnotetext[1]{Some of this material was presented during the 4th
General Assembly of EGU 2008 at the session NH1.02 titled ``Extreme
Events Induced by Weather and Climate Change: Evaluation, Forecasting
and Proactive Planning.''}

\footnotetext[2]{Atmospheric and Environmental Research, Inc.,
Lexington, Massachusetts, USA.}

\footnotetext[3]{\em Contact information:
\upshape
Dr. Ross N. Hoffman,
Atmospheric and Environmental Research, Inc.,
131 Hartwell Avenue,
Lexington, MA 02421-3126
Email: ross.n.hoffman@aer.com.}

\maketitle

\pagestyle{myheadings}

\markboth{Hoffman \etal}{Potential of exigent forecasting of severe weather}

\begin{abstract}
Severe storms, tropical cyclones, and associated tornadoes, floods,
lightning, and microbursts threaten life and property.
Reliable, precise, and accurate alerts of these phenomena can trigger
defensive actions and preparations.
However, these crucial weather phenomena are difficult to forecast.
The objective of this paper is to demonstrate the potential of \dvar4
(four-dimensional variational data assimilation) for exigent
forecasting (XF) of severe storm precursors and to thereby
characterize the probability of a worst-case scenario.
\dvar4 is designed to adjust the initial conditions (IC) of a
numerical weather prediction model consistent with the uncertainty of
the prior estimate of the IC while at the same time minimizing the
misfit to available observations.
For XF, the same approach is taken but instead of fitting observations,
a measure of damage or loss or an equivalent proxy is maximized or
minimized.
For example, XF of maximized significant tornado parameter (STP) would
delineate relative probabilities of the threat of tornadogenesis as a
function of time and place.
To accomplish this will require development of a specialized cost
function for \dvar4.
When \dvar4 solves the XF problem a by-product will be the value of
the background cost function that provides a measure of the likelihood
of occurrence of the forecast exigent conditions, and the value of the
STP cost function that provides an estimate of the likelihood of
tornadogenesis.
\dvar4 has been previously applied to a special case of XF in
hurricane modification research.
A summary of a case study of Hurricane Andrew (1992) is presented as a
prototype of XF.

The study of XF is expected to advance forecasting high impact weather
events, refine methodologies for communicating warning and
potential impacts of exigent weather events to a threatened
population, be extensible to commercially viable products, such as
forecasting freezes for the citrus industry, and be a useful
pedagogical tool.
Further, by including parameter sensitivity in the adjoint model, XF
could be extended to include parametric uncertainty.
\end{abstract}

\clearpage

\twocolumn

\xx {Introduction}

Forecasting high impact weather associated with severe thunderstorms and
tropical cyclones (TCs) is a challenging intellectual problem that has
the potential to save lives and protect property.
High impact weather associated with severe storms and TCs includes
lightning, hail, floods, tornadoes (\figr{tornado}), and microbursts.
Currently, such small-scale weather phenomena can not be usefully
forecast by explicit dynamic numerical weather prediction (NWP)
methods.
NWP solves the initial value problem for the fluid dynamical equations
that govern the evolution of the atmosphere.
In applications of NWP methods, data assimilation systems
account for the uncertainties present in the models due to errors in
the model representation of the atmosphere and in the initial
conditions (IC) due to errors and gaps in the observations of the
atmospheric parameters.
While NWP model representation of the fine scale structure of the
atmosphere is steadily improving and today's high resolution research
models can predict realistic severe storms and TCs, predictions of
precise timing and location and details of internal storm dynamics do
not agree well with reality.
For example,
forecasts of actual tornadoes would require ultra-high resolution that
is not yet feasible, as well as more capable observation networks.
Instead, useful tornado forecasts focus on prediction of the
larger-scale characteristics of the environment of severe
storms---large thermodynamic instability, vertical wind shear, etc.
Such empirical forecasts are applicable and reliable to the extent
that it is the large-scale environment that regulates the small-scale
high impact weather phenomena.

%%%%%%%%%%%%%%%%%%%%%%%%%%%%%%%%%%%%%%%%%%%%%%%%%%%%%%%%%%%%%%%%%%%%%%%%%%%%%%

 \fitem{tornado}{The Greensburg, KS Tornado.  Greensburg, KS was
 destroyed on 4 May 2007.
 Photo by Van DeWald, NWS.
 From www.hprcc.unl.edu/nebraska/may4-2007Greensburg-Kansas-tornado.html.}
{\includegraphics[width=3in,bb=0 0 800 475]{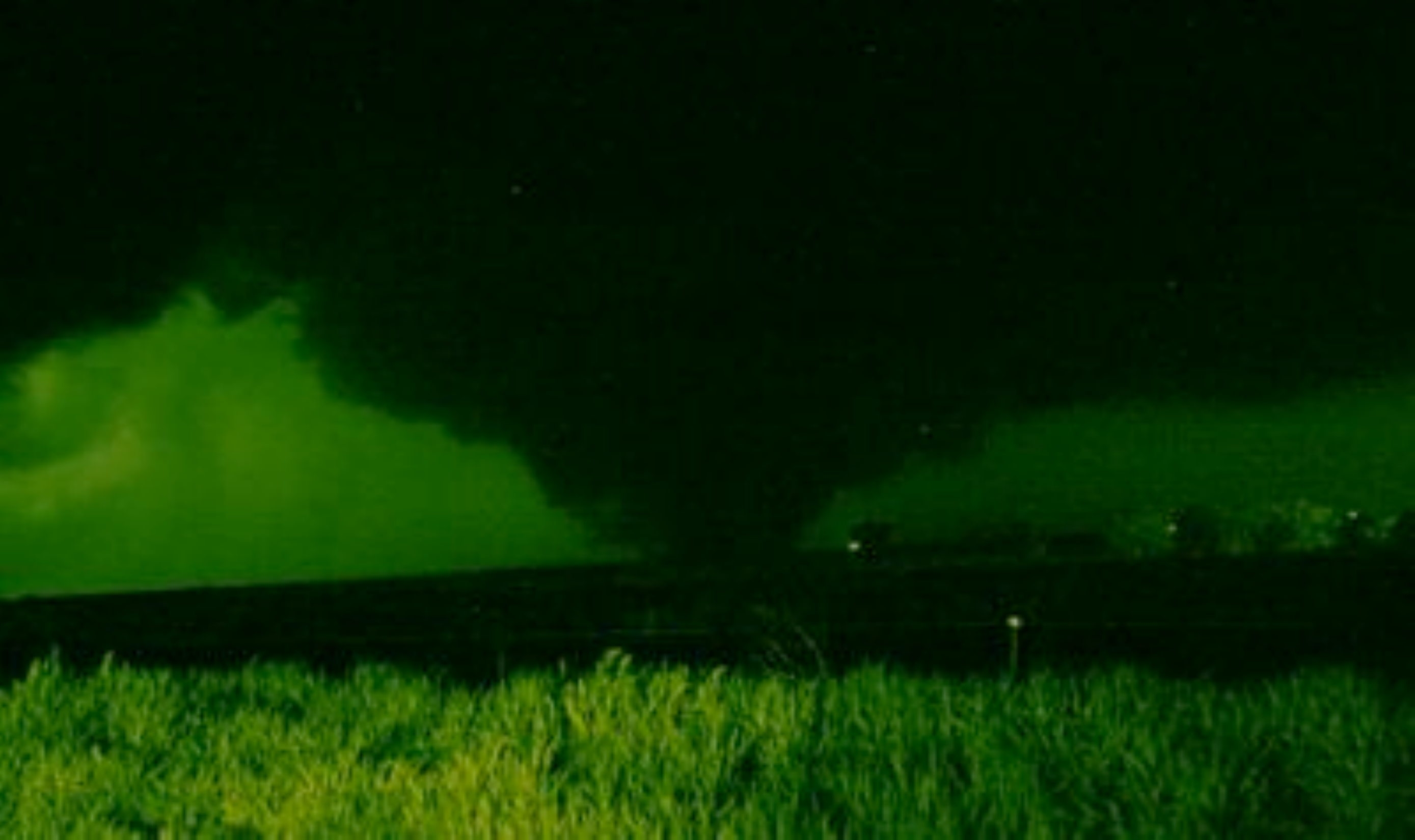}}

%%%%%%%%%%%%%%%%%%%%%%%%%%%%%%%%%%%%%%%%%%%%%%%%%%%%%%%%%%%%%%%%%%%%%%%%%%%%%%

A characteristic of models of the atmosphere is extreme
sensitivity to small perturbations of the IC.
This sensitivity, coupled with the uncertainty inherent in models and
observations, leads to a loss of predictability.
Consequently, it is possible that a small change in the IC could
produce a very different forecast.
That is, a forecast of a very large specific impact on people,
property, or nature might result from a particular (small,
plausible) IC perturbation, \ie, one that is consistent with the
uncertainties present.
It turns out that it is possible, using a technique called \dvar4
described below, to calculate the IC perturbation to produce a
particular result.
For example, suppose we calculate the perturbations to maximize the
potential for tornadogensis.
Then, if the calculated IC perturbations are small compared to the IC
uncertainty, we must not discount the possibility of tornadoes
forming even if the original (\ie, unperturbed) forecast did not
indicate this possibility.
\citet{HenHL+05} call this approach ``exigent'' forecasting because of
the requirement of precision in the forecast and urgency in the
response.
In this paper we describe the potential application of exigent
forecasting (XF) to forecasting severe weather precursors for
tornadoes (\secr{application}).
Because of current limitations for forecasting very small scales, such
as severe storms and tornadoes, applications of XF must first focus on
larger scales, such as the strength or path of a TC, or on the
precursors to the small-scale high impact phenomena.
In previous work we applied a slight variation of the XF approach to
controlling TCs and a summary of that work is provided to demonstrate
that XF works (\secr{tcmod}).

Our implementation of XF is based on \dvar4 (four-dimensional variational data
assimilation).
\dvar4 is designed to adjust the IC of a numerical weather prediction
model consistent with the uncertainty of the prior estimate of the IC
while at the same time minimizing the misfit to available
observations.
For XF, the same approach is taken but instead of (or in addition to)
fitting observations, a measure of damage, loss, or an equivalent
proxy is maximized or minimized (\secr{method}).
For example, XF of maximized significant tornado parameter (STP,
\secr[~\ref{sec:application}.]{STP}) should be able to provide relative probabilities of the
threat of tornadogenesis as a function of time and place.
To accomplish this, a specialized cost function for \dvar4 is needed.
This cost function is designed to measure the likelihood and severity of
tornado outbreaks, based on the synoptic and
mesoscale conditions conducive to their formation---the cost
function does not measure tornado damage directly.
That is, we use the STP as a proxy that captures the key environmental
controls of tornadic development.
When \dvar4 solves the XF problem a by-product will be the value of
the background cost function that provides a measure of the likelihood
of occurrence of the forecast exigent conditions, and the value of the
STP cost function that provides an estimate of the likelihood of
tornadogenesis.
XF could be a useful complement to ensemble forecasting because the
ensemble is designed to capture the overall distribution of future
outcomes, but not necessarily the particular event of consequence that
is the focus of XF.

In the future, XF is expected to have very significant potential for
societal benefits and to be applicable to a wide array of problems (\secr{future}).
We note that operational forecasting of severe high-impact weather and
subsequent decision-making is extremely time-sensitive.
We comment on efficient implementation briefly in \secr[~\ref{sec:application}.]{mode}.
XF has the potential to provide specific guidance of worst case
results to focus the forecast and decision-making activity.
The study of XF is expected to advance knowledge of forecasting high
impact weather events and to result in refined methodologies for
communicating warning and potential impacts of exigent weather events
on a threatened population.
XF might be extended to commercially viable products, such as
forecasting freezes for the citrus industry.
In addition, XF could be extended to deal with parametric uncertainty
(\eg, the uncertain knowledge of the surface roughness height, $z_0$).
Finally, XF should prove to be a useful device for laboratory exercises
at the university level.

\xx {\dvar4 methodology\secl{method}}

For operational weather forecasting, \dvar4 finds the smallest
increment at the start of each data assimilation period so that the
perturbed nonlinear solution best fits all the available data.
Mathematically, \dvar4 blends background (\ie, a short-term forecast)
and observations by minimizing a functional, \eql{Jbo}{J = \J{b} +
\J{o},} with respect to a control vector \Vector{x} that describes the
atmospheric state.
Here \J{b} and \J{o} measure the misfit of the four-dimensional analysis
(\ie, the simulation that begins with \Vector{x}) to the background
and observations, respectively.
\dvar4 solves this complex nonlinear minimization problem iteratively,
making use of the linear adjoint of the model, linearized about the
current nonlinear simulation.
Note that \J{b} is defined by specifying the forecast error
covariance---an unresolved problem for the storm-scale as well as for
TCs and other severe weather.
For exploratory experiments, available climatological background error
covariances appropriate for mesoscale forecasts of similar resolution
and for similar synoptic regimes could be used.
A hybrid ensemble/\dvar4 would provide improved covariances
(\secr[~\ref{sec:future}.]{beyond}).

\dvar4 and XF are useful and indeed possible because the atmosphere is
chaotic and unpredictable.
Atmospheric motions occur over a huge spectrum of scales ranging from
the jet stream to the trajectory of a cloud drop.
Errors at different scales are coupled by nonlinear interactions.
Tiny errors inevitably present in the large scales
very quickly result in errors in the position of small-scale features.
At the same time, errors present in small scales grow quickly and
soon become unpredictable.
These errors then effect somewhat larger scales and so on.
The net effect is that very small perturbations can grow quickly.
\dvar4 finds perturbations that grow in just the correct way to best
fit the available data or to best satisfy the XF constraints. 

A few changes must be made to \dvar4 for XF \citep{HenHL+05}: the
control vector remains \Vector{x}, \J{b} is unchanged, \J{o} may be
present or not depending on whether more recent observations are
available, and \J{d} (``d'' for ``damage'') is added to measure the
``benefits'' minus ``costs'' resulting from the forecast as defined by
the customer.
The functional to be minimized for XF is then \eql{Jboc}{J = \J{b} +
\J{o} + w_d \J{d}.}
The definition of \J{d} should be nondimensional and $w_d$ is an
adjustable weight.
As $w_d$ is increased, the value of \J{d} determined will decrease and
vice versa.
Note that \J{d} is defined to calculate a worst case scenario:
minimizing \J{} will simultaneously maximize costs, minimize benefits,
and minimize increments \wrt\ the background and observations.
By contrast, in our hurricane modification experiments (\secr{tcmod}),
by defining \J{d} to measure damage alone, our solutions must minimize
damage.
That is, we calculated a ``best case scenario'' to determine
perturbations to reduce damage.

In \dvar4 data assimilation, the final value of $J_b + J_o$
objectively quantifies the likelihood of the calculated changes to
\Vector{x}, based on the most recent background, the
observations, and knowledge of instrument and model error.
This is the case since the log of the probability of an atmospheric
state, given the background, the observations, and the associated
uncertainties, is directly related to $J_b + J_o$ according to Bayes'
theorem when suitable assumptions are valid \citep{Lor86}.
Note that smaller values of \J{} indicate higher probability.
Since this applies to any atmospheric state, the final value of \J{b}
in the XF case will still be related to
the likelihood of the exigent solution occuring.
It is expected that \J{} has a $\chi^2$-distribution
\citep{DesI01,MucHB04}.
In the XF situation, the final value of \J{b} will still be related to
the likelihood of the exigent solution occuring.
As described in the next paragraph, by calculating the final value of
\J{b} as a function of the region used in the definition of \J{d}, we
will obtain a map of the relative probability of the exigent event
(\eg, of the occurrence of tornadogenesis).

Ideally the objective evaluation would be combined with a subjective
evaluation based on forecaster experience---including pattern
matching---that would compare the new analysis and observations to
determine if the perturbation is consistent with hypothetical, but
plausible, dynamical processes or additional observations.
A series of solutions with increasing $w_d$, the weight given to
\J{d}, gives a series of solutions that are increasingly unlikely
(\ie, larger \J{b}) and associated with increasing (tornado) threat.
One would stop increasing $w_d$ when the subjective evaluation
indicates a nonrealistic solution or when the calculated likelihood
becomes smaller than a prespecified lower value.
Plotting the values, respectively, of \J{d} and \J{b} at the minimizing
solution, with respect to $w_d$ for each location and forecast time,
will allow us to highlight the locations and times of greatest threat.

%%%%%%%%%%%%%%%%%%%%%%%%%%%%%%%%%%%%%%%%%%%%%%%%%%%%%%%%%%%%%%%%%%%%%%%%%%%%%%

 \fitem{track} {Forecast tracks of Hurricane Iniki.
The center of the simulated hurricane is plotted for each hour for
\hr{t_0 = 0} to \hr{24} for the unperturbed simulation (black dots)
and for two target experiments, one allowing perturbations to all
variables (orange) and one allowing only temperature perturbations
(green).
For reference, these tracks are plotted over the wind speeds of the
repositioned cyclone used as the goal.
Wind speeds corresponding to different Saffir-Simpson categories are
plotted with different colors: grey for tropical depressions (\ms{>
12}), green for tropical storms (\ms{> 17}), yellow for category 1
hurricanes (\ms{> 33}) and red for category 2 hurricanes (\ms{> 43}).
After \citet{HofHL+06b}.}
{\fbox{\includegraphics[scale=1, bb=200 325 400 525, clip=true]{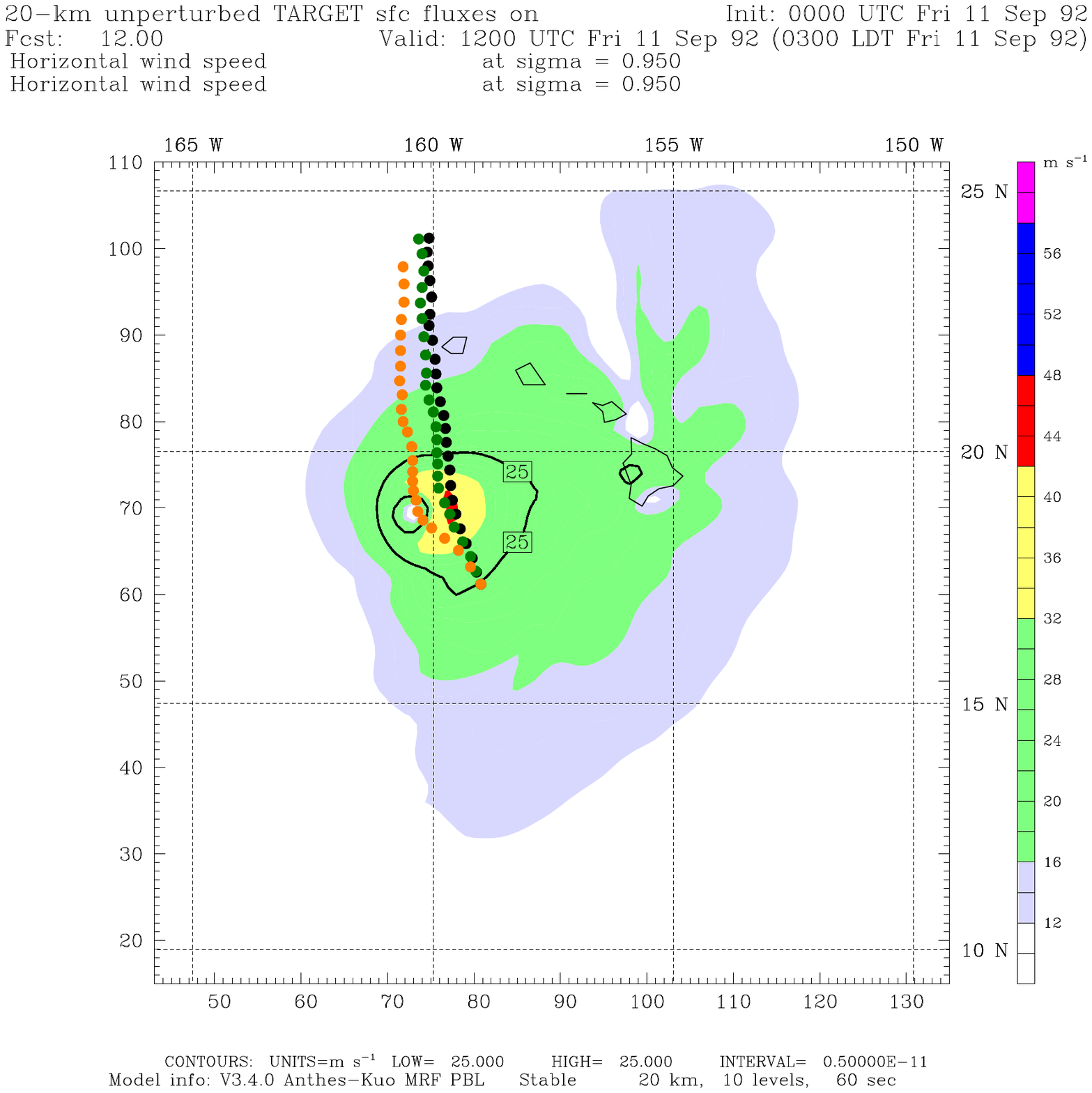}}}

%%%%%%%%%%%%%%%%%%%%%%%%%%%%%%%%%%%%%%%%%%%%%%%%%%%%%%%%%%%%%%%%%%%%%%%%%%%%%%

\xx {An XF prototype: Weather control for hurricanes\secl{tcmod}}

Experiments we conducted demonstrate the ability of \dvar4 to
calculate the influence of perturbation on the future path or
intensity of a simulated TC.
In ``target'' experiments, the MM5 \dvar4 system determines the
optimal atmospheric state trajectory, which simultaneously minimizes
the size of the initial perturbation and the difference (using a
quadratic norm) between the new forecast state and a goal atmospheric
state in which the simulated cyclone has been repositioned.
That is, \J{d} in these experiments is similar in form to \J{b}, but
instead of measuring the difference between the analysis and the
background, \J{d} here measures the difference between a short (in the
case presented later six-hour) forecast and the goal or target.
In experiments for Hurricane Iniki, the simulated hurricane
successfully matched the goal state at the end of the \dvar4 period,
then continued on a track parallel to its original track, and missed
the Island of Kauai as desired \citep{HofHL+06b}.
Perturbations to the initial conditions are small relative to the
hurricane.
When changes are restricted to the temperature field alone, the
technique is less successful (\figr{track}).

In ``damage cost function'' experiments, \dvar4 simultaneously
minimized the size of the initial perturbation and an estimate of
property loss that depends on wind speed.
In these experiments \J{d} is given by \eql{Jd}{J_d =
\frac{1}{\Delta t}\frac{1}{A}\int\limits_{\Delta t}\int\limits_A
D(x,y,t) P(x,y),} where $\Delta t$ is the time period
of interest and $A$ is the area of interest.
The fractional wind damage $D$ is taken to depend only on the lowest
model layer wind speed at location $(x,y)$ with the functional form
shown in \figr{windJd}, and the property values $P(x,y)$ were assigned
based on a land use data base.
(Details are given by \citet{HenHL+05}.)

%%%%%%%%%%%%%%%%%%%%%%%%%%%%%%%%%%%%%%%%%%%%%%%%%%%%%%%%%%%%%%%%%%%%%%%%%%%%%%

\fitem{windJd} {Wind cost function.
In our model of wind damage, wind speeds less than \ms{25} cause no
damage and wind speeds in excess of \ms{90} cause total loss.
Between these two threshold wind speeds a cosine curve parameterizes
the percent of wind damage.}
{\includegraphics[scale=0.4,bb=0 0 495 366]{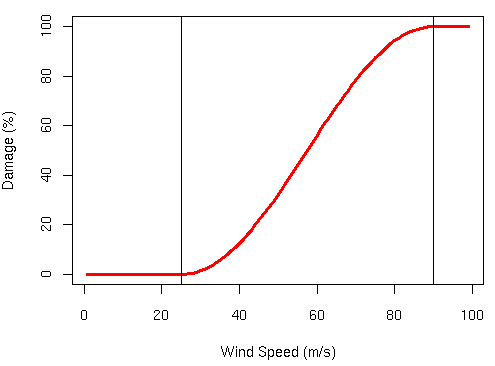}}

%%%%%%%%%%%%%%%%%%%%%%%%%%%%%%%%%%%%%%%%%%%%%%%%%%%%%%%%%%%%%%%%%%%%%%%%%%%%%%

In experiments for Hurricane Andrew, the hurricane surface winds
decrease over the built-up area at landfall \citep{HenHL+05}.
Additional experiments explored the ability of other state variables
to effect desired changes to Hurricane Iniki.
Perturbations restricted to winds
alone, to temperature alone, or even to temperature only outside
the center of the hurricane were found to be effective
\citep{HofHL+06a}.
Furthermore, vertical velocity and humidity perturbations alone were ineffective at
reducing damaging winds.
An experiment using perturbation pressure alone substantially
reduced the extent of damaging winds. While this result is similar to
experiments that used solely temperature in the control vector, the
minimization failed to converge to within specified numerical limits
and thus should be considered less robust.

The optimal perturbations usually include quasi-axisymmetric features
centered on the hurricane (\figr{Andrew3dTIC}).
It appears that the perturbations then evolve as concentric wave
disturbances that propagate to a focus at the hurricane center, and
convert the kinetic energy of the hurricane into thermal potential
energy at the appropriate time.
The hurricane surface winds regenerate soon thereafter, so a
continuous series of perturbations may be needed in practice
(\figr{windsLT}).

%%%%%%%%%%%%%%%%%%%%%%%%%%%%%%%%%%%%%%%%%%%%%%%%%%%%%%%%%%%%%%%%%%%%%%%%%%%%%%

\fitem{Andrew3dTIC} {Initial perturbation for Hurricane Andrew.
In this color volume rendering of optimal initial temperature
perturbations for Hurricane Andrew, the red and blue surfaces enclose
the volumes where the perturbations are greater than \degreesC{0.6},
and less than \degreesC{-0.6}, respectively.}
{\includegraphics[scale=0.3,bb=0 0 711 761]{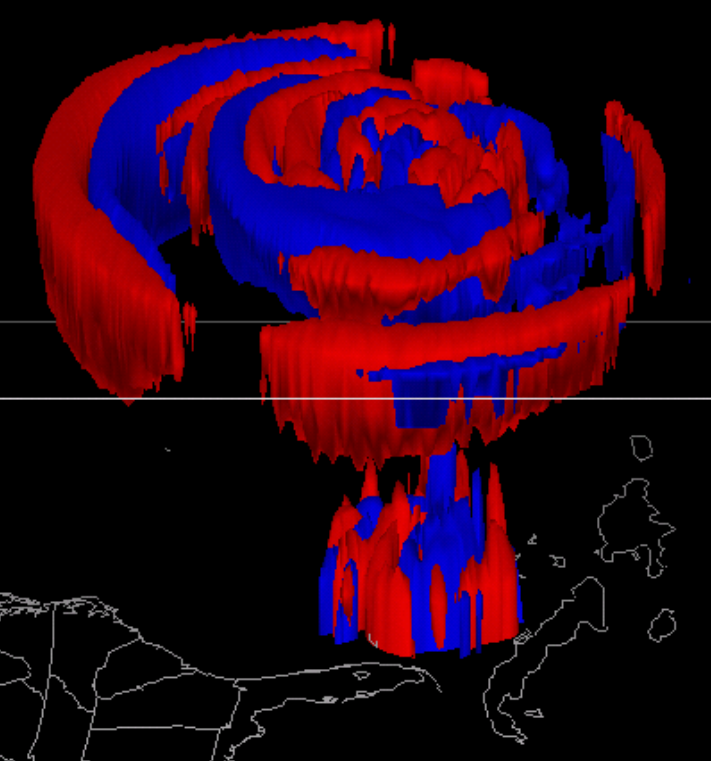}}

%%%%%%%%%%%%%%%%%%%%%%%%%%%%%%%%%%%%%%%%%%%%%%%%%%%%%%%%%%%%%%%%%%%%%%%%%%%%%%

 \fitemstar{windsLT}{Hurricane Andrew surface wind fields.
Evolution of the surface wind field for the unperturbed simulation and
for the controlled experiment that minimized damage between 4 and
\hr{6} by introducing temperature perturbations only.
Wind speed is color coded according to the Saffir-Simpson scale (as in
\figr{track}) and the damaging wind contour \ms{25} is plotted at 4, 6,
and \hrs8 after the perturbation is introduced.
In the experiments only wind speeds above \ms{25} result in property
damage.
Note that at \hr{6} there are no damaging winds over land areas in the
controlled experiment.
After \citet{HenHL+05}.}
{\setlength{\tabcolsep}{1pt}
 \begin{tabular}{|c|c|c|c|}
 \hline
 \makebox[8mm]{\rule[-3mm]{0mm}{8mm}} &
 4 h & 6 h & 8 h \\ \hline
 \ylab{Unperturbed} & 
 \tfbox{\includegraphics[scale=0.73, bb=125 275 325 475, clip=true]{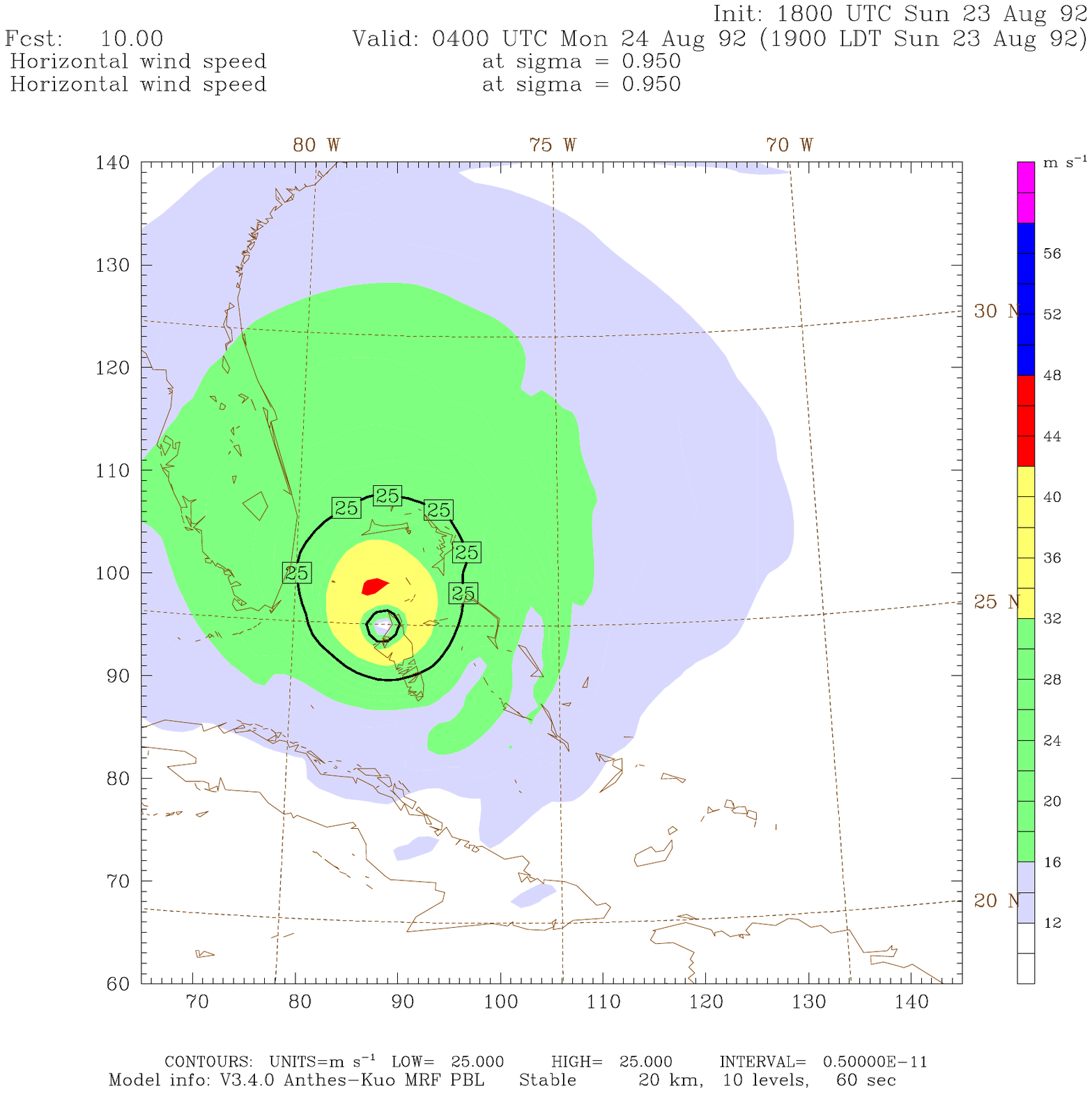}} &
 \tfbox{\includegraphics[scale=0.73, bb=125 275 325 475, clip=true]{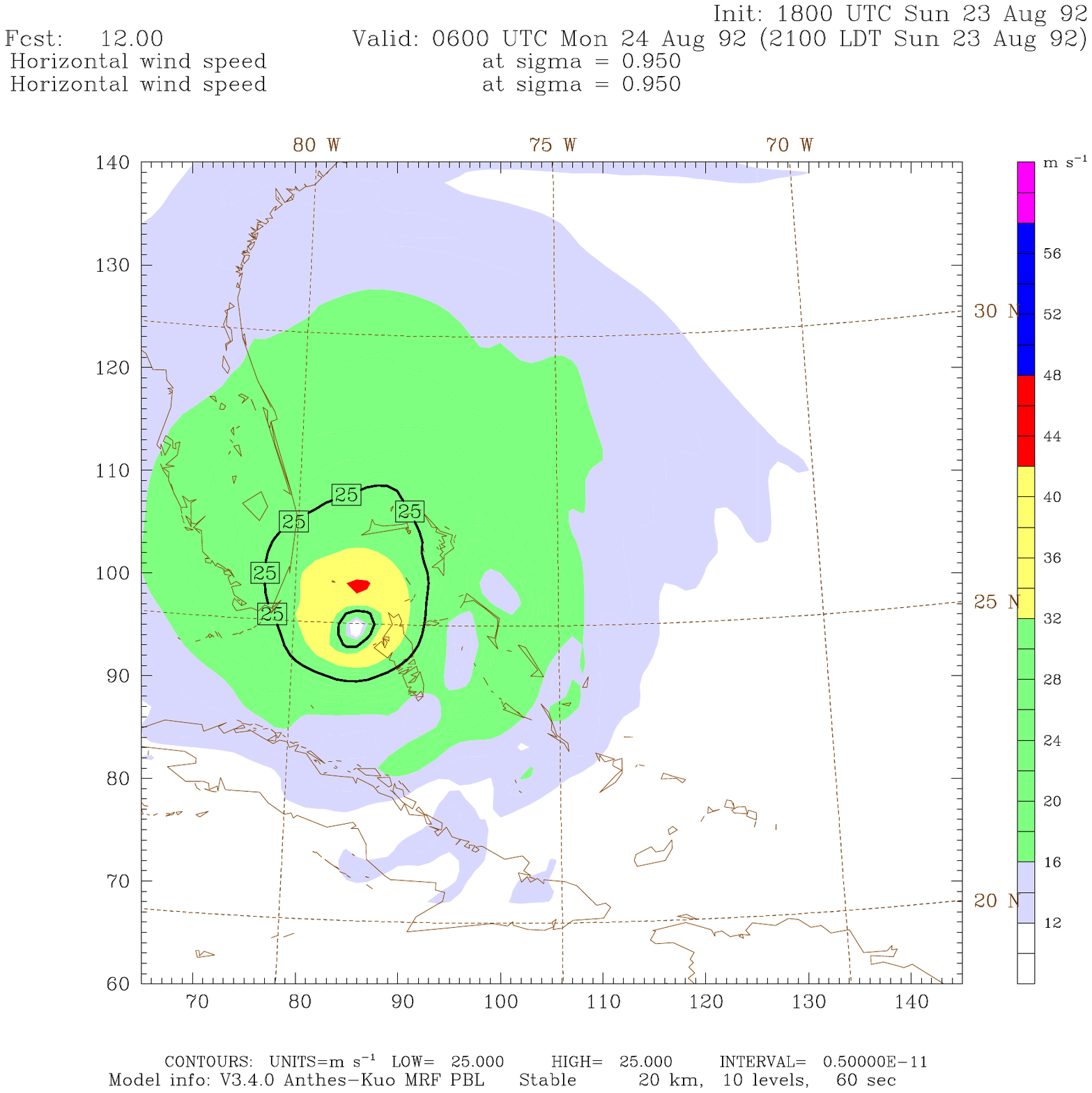}} &
 \tfbox{\includegraphics[scale=0.73, bb=125 275 325 475, clip=true]{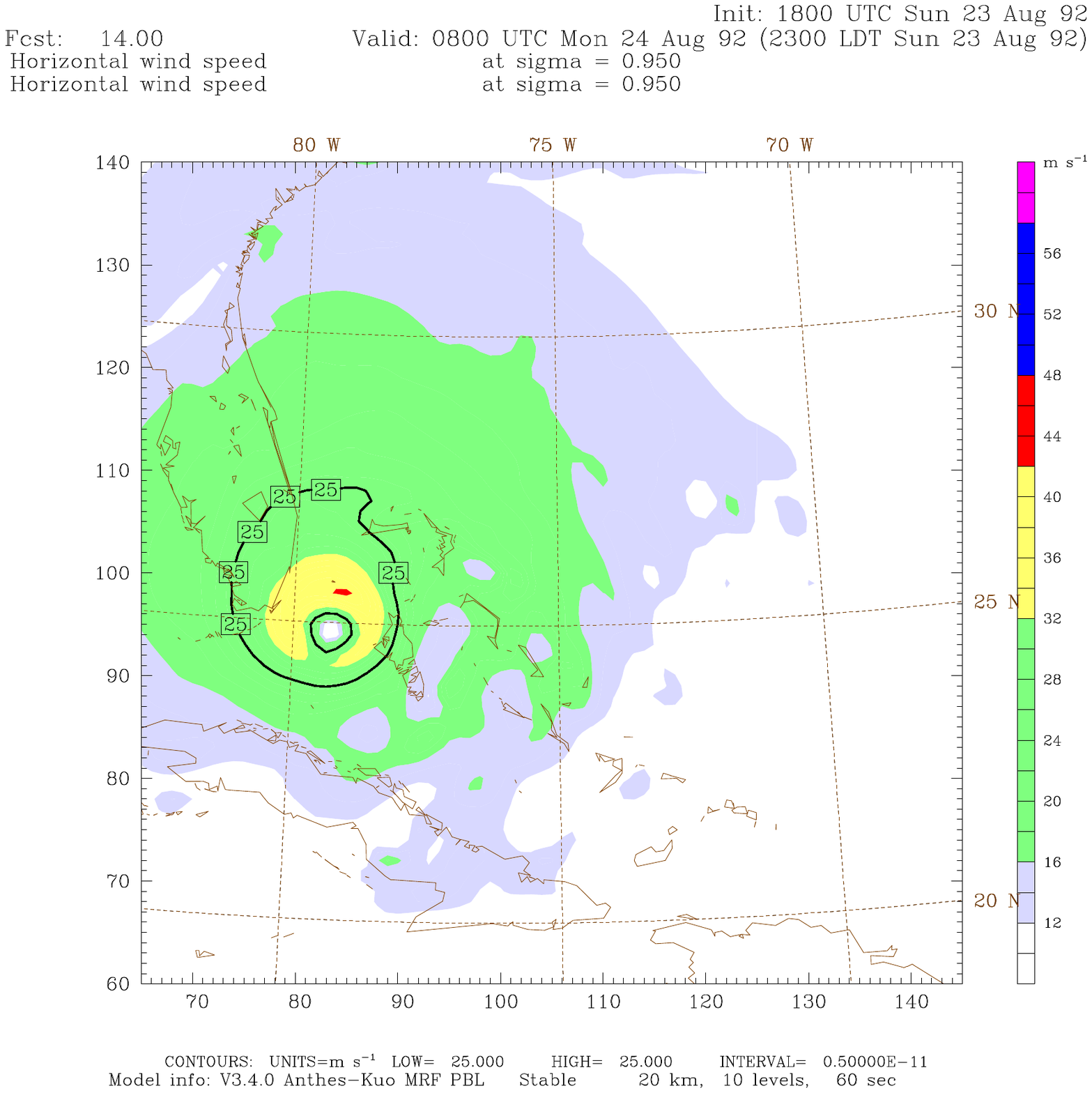}} \\ \hline
 \ylab{Controlled} &
 \tfbox{\includegraphics[scale=0.73, bb=125 275 325 475, clip=true]{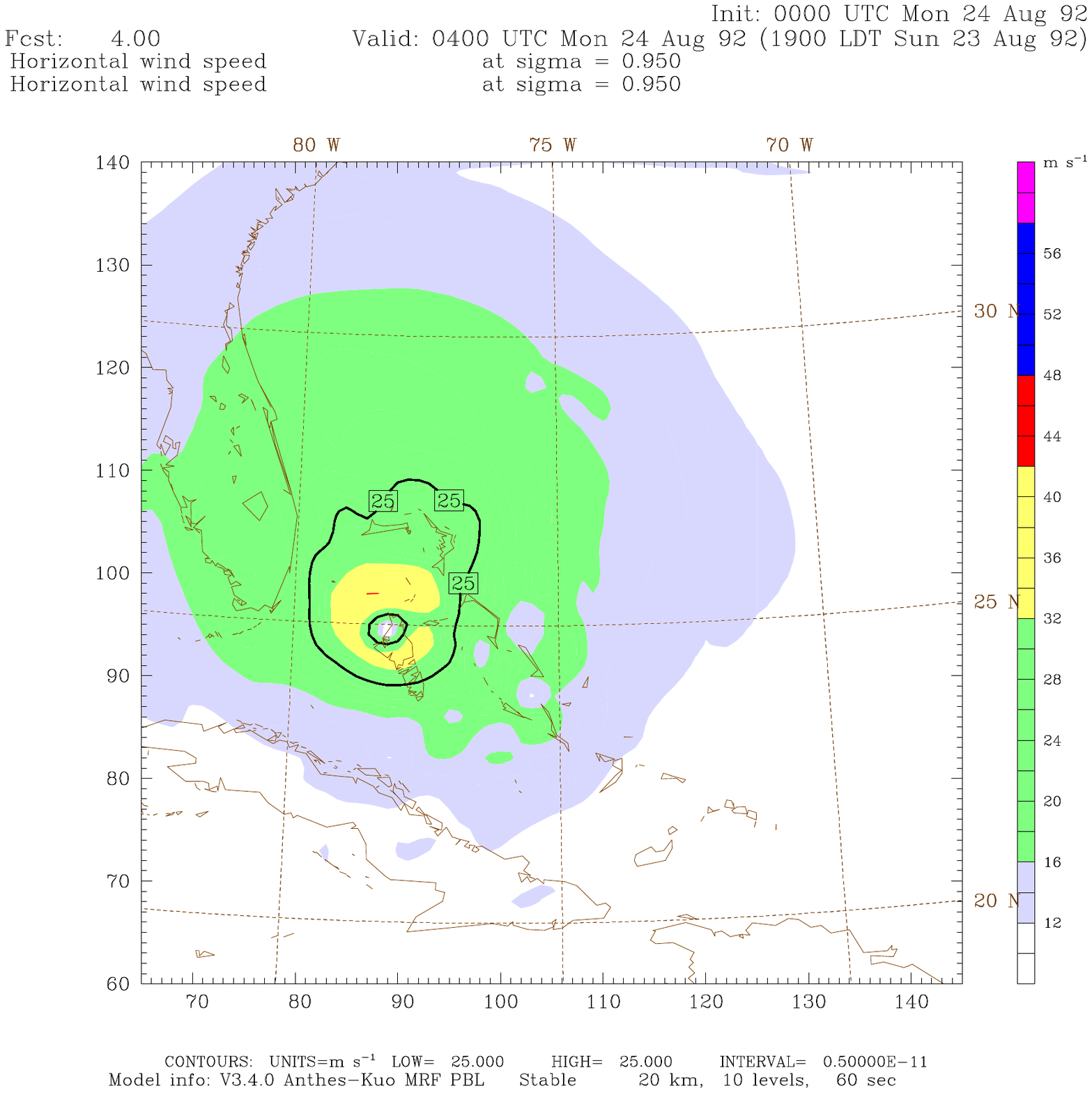}} &
 \tfbox{\includegraphics[scale=0.73, bb=125 275 325 475, clip=true]{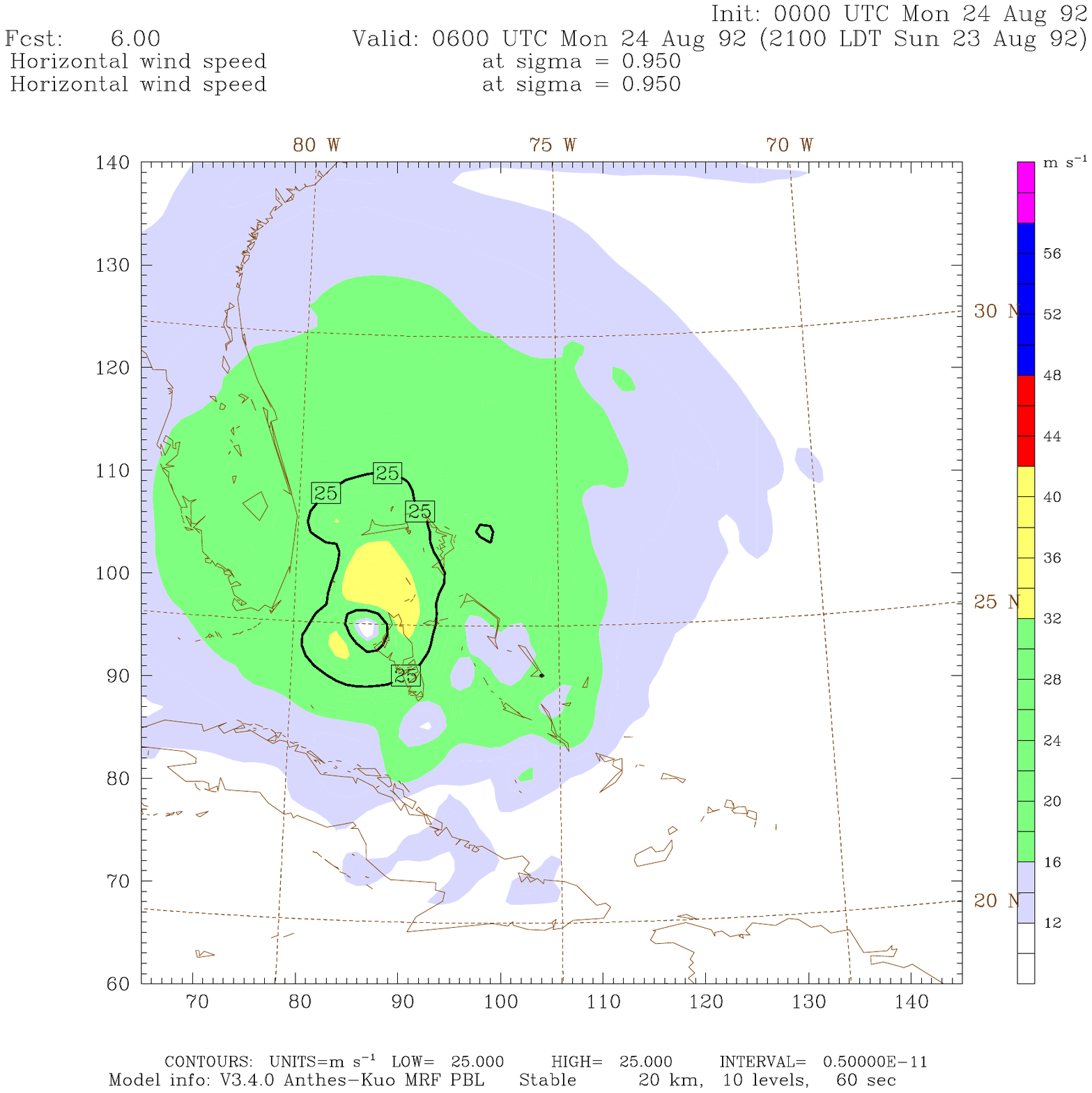}} &
 \tfbox{\includegraphics[scale=0.73, bb=125 275 325 475, clip=true]{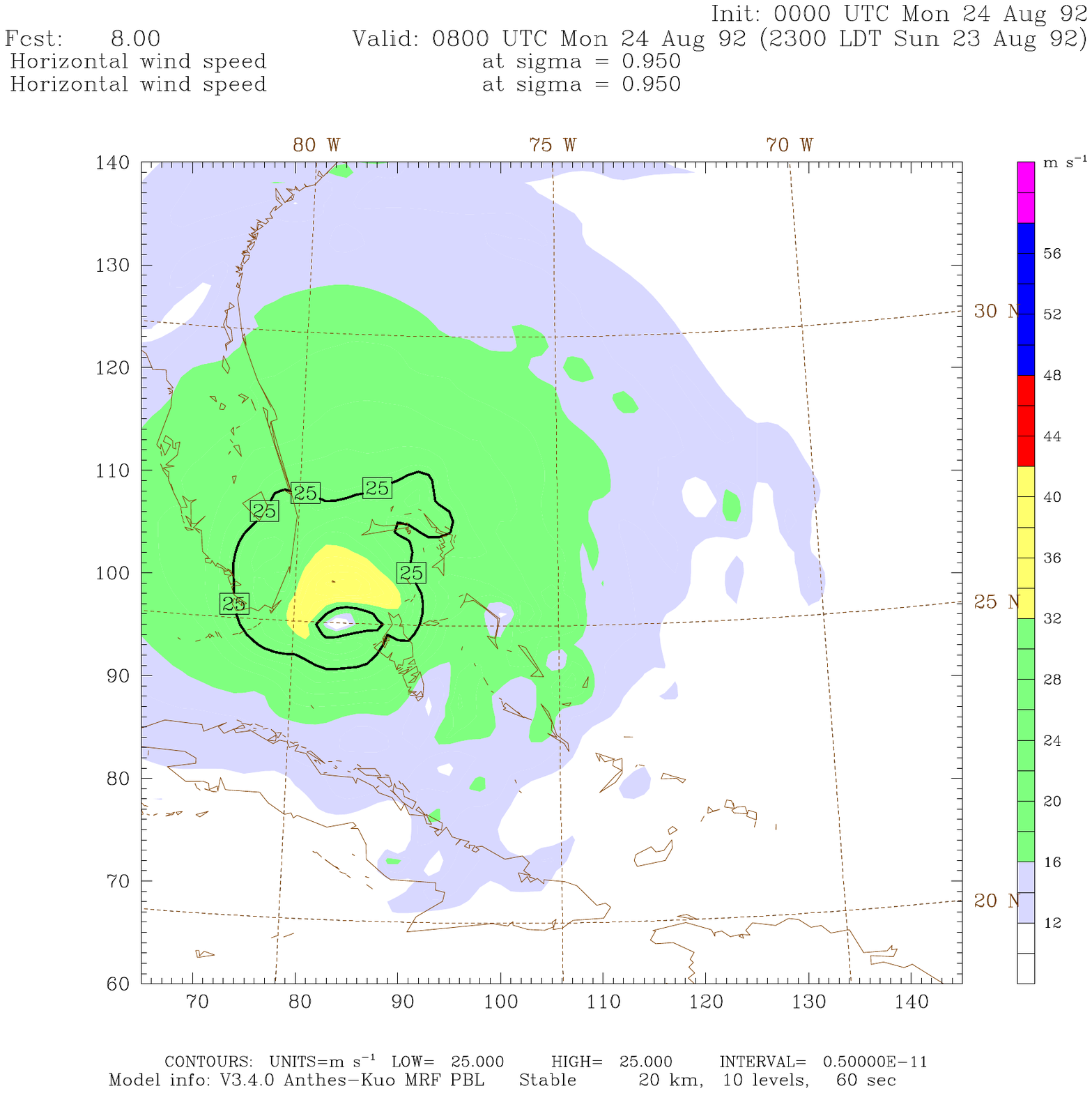}} \\ \hline
 \end{tabular}}

%%%%%%%%%%%%%%%%%%%%%%%%%%%%%%%%%%%%%%%%%%%%%%%%%%%%%%%%%%%%%%%%%%%%%%%%%%%%%%

\clearpage

\xx {Application to  tornadogenesis precursors\secl{application}}

A combination of inaccuracy in the initial conditions (IC), model
errors, and the short predictability time scale for severe storms all
combine to make tornado forecasting very difficult.
Especially challenging are parameterizing the moist microphysics and
estimating the IC for water vapor, cloud, and precipitation since in
stormy environments these have small spatial scales
\citep[\eg,][]{WeiKB+07}.
As a result,
prediction of severe weather generally uses an ingredients-based
forecast method wherein coincidence of multiple individual atmospheric
predictors increases forecaster confidence in the likelihood of
severe weather.
Historically, the risk of severe weather is maximized when and where
the ingredients coincide.
The discussion here will focus on one aspect of this overall
problem---the IC uncertainty for dry-line tornadogenesis, where the
empirical ingredients-based approaches work best.
The XF approach has greater applicability than this one focus.
In other parts of this paper, we sketch the extensions needed to handle
directly forecasting storm scale high impact weather elements as well
as other high impact or high value weather events such as landfalling
hurricanes and freezing conditions in citrus growing regions.

Ingredients-based indices, due to their nature, tend to focus on
prediction of sensible weather conditions, such as the likelihood of
supercell thunderstorms or significant tornadoes, whereas individual
indices or ingredients describe more esoteric atmospheric features,
such as the magnitude of thermal instability.

\xxx {Cost function for dryline tornadogenesis\secl{STP}}

XF of tornadogenesis based on \dvar4 requires a cost function based on
an ingredients-based index.
One ingredients-based index that can define \J{d} is the
significant tornado parameter (hereafter, STP).
The STP combines individual measures of atmospheric thermal stability,
boundary layer moisture and horizontal wind fields.
This index has been shown subjectively to discriminate between
convective events \emph{involving discrete supercells} that produce
significant tornadoes and those that produce weak tornadoes or none at
all.
The requirement that the mode of the expected convection be discrete
supercells limits the applicability of STP-based XF.
In other situations, other more appropriate indices should be used.

As defined by \citet{ThoEH+03},
\begin{eqnarray} \label{eq:STP} 
\mbox{STP} &=& 
\frac{\mbox{MLCAPE}}{\mks{1000}{J\;kg^{-1}}} \:\times\:
\frac{\mbox{SHR}}{\ms{20}}
\nonumber \\ & \times & 
\frac{\mbox{SRH}}{\mks{100}{m^2\;s^{-2}}} \:\times\:
\frac{2000-\mbox{MLLCL}}{\mks{1500}{m}}
\nonumber \\ & \times &
\frac{150-\mbox{MLCIN}}{\mks{125}{J\;kg^{-1}}},
\end{eqnarray}
where MLCAPE is the mixed-layer convective available potential energy
(CAPE), SHR is the magnitude of the 0-6 km vertical shear of the
horizontal wind, SRH is the magnitude of the 0-1 km storm relative
helicity, MLLCL is the mixed-layer lifted condensation level (LCL),
and MLCIN is the mixed-layer convective inhibition (CIN).
Values of STP greater than one are associated with the potential for
significant tornadoes.
Each of the components of STP has a history of use in tornado
prediction and each describes characteristics of the local storm
environment as detailed in \tabr{stp}.
We believe that use of the STP as a penalty function may be especially
relevant to characterizing the likelihood of CIN erosion---a frequent
forecasting problem and one that strongly modulates the overall extent
of convection.
The definition of STP is empirical and continues to be refined by
\citet{ThoEM04,ThoME07}.
However, for a preliminary study the definition given in
\eqr{STP} should be adequate.
A straightforward approach to defining \J{d} so that minimizing $J$
defined in \eqr{Jboc} will result in maximizing the STP is to average
STP over a target area $A$, and time interval $\Delta t$:
\eql{JD1}{J_d = -\frac{1}{\Delta t}\frac{1}{A}\int\limits_{\Delta
t}\int\limits_A\mbox{STP}.}
Typically $A$ will be a county of interest, and $\Delta t$ a period of
interest lasting from one to several hours.

%%%%%%%%%%%%%%%%%%%%%%%%%%%%%%%%%%%%%%%%%%%%%%%%%%%%%%%%%%%%%%%%%%%%%%%%%%%%%%

\begin{table*}[t] \caption{STP ingredients\label{tab:stp}}
\vspace*{4mm}  \centering
\begin{tabular}{cc}
\tophline

\ylab{\bf CAPE} & \tpbox{CAPE is a measure of the instability through
the depth of the atmosphere, \ie, the positive buoyant energy
available, and is generally proportional to the updraft strength of a
storm.
Values of CAPE over \mks{4000}{kg^{-1}} indicate extreme instability
likely associated with high lapse rates.} \\ \middlehline

\ylab{\bf SHR} & \tpbox{The magnitude of the deep-layer wind shear,
along with the amount of instability, regulates the overall dynamical
organization and persistence of a storm.
Values of wind shear greater than \ms{20} in NCEP Rapid Update Cycle
Version 2 model soundings have been shown by \citet{ThoEH+03} to
strongly support development of supercells.}\\ \middlehline

\ylab{\bf SRH} & \tpbox{The storm-relative helicity is a measure of
the potential of low-level cyclonic updraft rotation for right-moving
supercells, \ie, supercells that move slightly to the right of the
mean flow for a cyclonically curved hodograph.
Values over \mks{100}{m^{2}\;s^{-2}} are often seen with tornadic
supercells.
\citet{WeiR00} describe how wind shear and helicity determine supercell
dynamics.}\\ \middlehline

\ylab{\bf MLLCL} & \tpbox{Lower values of environmental LCL, related
to high mixed-layer relative humidities, tend to increase the
potential buoyancy of rear flank downdrafts in storms capable of
supporting significant tornadoes \citep{MarSR03}.}\\ \middlehline

\ylab{\bf CIN} & \tpbox{CIN is a measure of the work required to lift
air to its level of free convection.
It represents the ``negative'' area on a sounding and is often likened
to a physical ``cap'' to explosive thunderstorm development.
A number of atmospheric processes can erode this negative area,
including synoptic scale upward motion, upward motion along a boundary
and surface heating.
Values of CIN associated with convection later in the day are often
greater (closer to zero) than -\mks{250}{J\;kg^{-1}}.
This term strongly regulates the spatial extent of significant STP.}\\
\bottomhline

\end{tabular}
\end{table*}

%%%%%%%%%%%%%%%%%%%%%%%%%%%%%%%%%%%%%%%%%%%%%%%%%%%%%%%%%%%%%%%%%%%%%%%%%%%%%%

As defined by \eqr{JD1}, \J{d} is a highly nonlinear function of the
atmospheric state vector, since not only is it a product of the five
separate terms in \eqr{STP}, but the individual terms themselves are
highly nonlinear: CAPE and CIN correspond to the areas between a
lifted parcel and the environmental sounding on a thermodynamic
diagram, and are highly sensitive to changes in the mixed layer
quantities that define the parcel.
Similarly, the storm-relative helicity depends on an estimate of storm
motion, which in turn depends on the depth of the convectively
unstable layer.
Formally, the nonlinearities in \J{d} can be treated in the same way
\dvar4 handles nonlinearities in model parameterizations and in
complex observation operators.
In the framework of iterative \dvar4, the full nonlinear formulation
is used in the outer loop, and linearized formulations (forward and
adjoint) are used in the inner loop \citep{Lor97,RabJK+00}.
However, strong nonlinearities, such as ``on-off'' switches from
stable to unstable conditions, can lead to convergence problems during
the minimization.
Therefore, to enhance the performance of \dvar4, some or all of the
individual terms of STP should be modified.
The modified versions will be better behaved in terms of being
differentiable and less nonlinear.
A general approach is to smooth out non-differentiable points.
Note that since STP has been defined and refined empirically,
relatively small changes to the formulation of STP should be
acceptable for current purposes.

\xxx {Key dryline tornadogenesis cases}

For demonstration, the \dvar4 XF method should be applied to 
Major Tornado Outbreak Days
\citep{SchBS04}, which have historically accounted for nearly 50\% of
all fatalities.
These are calendar days during which there are at least six
significant tornadoes, \ie, F2 and higher rating on the Fujita scale
\citep{Fuj71}.
(It should be noted that the Fujita scale is used here and not the
newer (February 2007) Enhanced Fujita Scale.)
Typically, Major Tornado Outbreak Days exhibit robust signatures for
tornadogenesis in soundings and NWP fields and are identified by SPC
forecasts as having a ``moderate'' or ``high'' risk for severe
weather.
Furthermore, initial case studies should be focused on
afternoon-evening springtime convective events in the form of discrete
supercells.
These conditions often are met near the dryline in the Central and
Southern Plains of the United States.
Finally, because small-scale boundaries in the domain caused by
previous convection (outflow boundaries, etc.) will not be well
represented by a model with currently available resolution and
parameterizations, only events which are substantially separate, in
space and time, from earlier convection should be considered.
Candidate demonstration cases are listed in \tabr{cases}.
The characteristics of the 1985 outbreak make it the archetypical
example for our purposes.

%%%%%%%%%%%%%%%%%%%%%%%%%%%%%%%%%%%%%%%%%%%%%%%%%%%%%%%%%%%%%%%%%%%%%%%%%%%%%%

\begin{table*}[t] \caption{Examples of dryline tornadogenesis\label{tab:cases}}
\vspace*{4mm}  \centering
\begin{tabular}{cc}
\tophline

\ylab{\bf 31 May 1985} & \tpbox{41 tornadoes, centered in NW
Pennsylvania and north of Toronto, Ontario, formed ahead of a cold
front during the afternoon and evening hours, killing 88 people.
The only F5 tornado in Pennsylvania history destroyed the town of
Wheatland.
\citet{For85} and \citet{FerOL86} have documented the statistics of the
outbreak, while \citet{FarC89} attributed an elevated mixed layer (EML)
as being a major factor in the generation of the large latent
instability available to the convection.
EMLs \citep{LanW91} are frequently associated with convective
outbreaks in the Plains of the US, yet rarely are displaced this far
to the northeast.
This case appears to be an excellent case to study due to its overall
robust and spatially distinct synoptic characteristics, predominant
supercellular mode of convection, and the removal of the ``cap'' below
the EML, \ie, there is no CIN below the EML.} \\ \middlehline

\ylab{\bf 3 May 1999} & \tpbox{This outbreak of 66 tornadoes in
Oklahoma and Kansas highlighted the risk to densely populated urban
areas when an F5 tornado passed through Moore, OK, a suburb of
Oklahoma City.
A total of 46 people were killed by the outbreak.
\citet{RoeSR02} and \citet{EdwCT+02} describe the subtleties in the
synoptic and mesocale environment of the outbreak which lowered
confidence in many aspects of the forecasting of this outbreak.} \\
\middlehline

\ylab{\bf 4-5 May 2007} & \tpbox{4-5 May 2007 - On back-to-back days
there were major tornado outbreaks in the Plains, including the first
EF5-rated tornado which destroyed Greensburg, KS \citep{McCRH07}.
Synoptic conditions were conducive to supercellular storms during both
afternoons and evenings in approximately the same geographical area.
On the 4th, great instability was present, but weak convergence along the
dryline, weak forcing aloft, and a capping inversion combined to 
limit the number of storms that formed,
especially farther south \citep{WeiKB+07}.} \\ \bottomhline

\end{tabular}
\end{table*}

%%%%%%%%%%%%%%%%%%%%%%%%%%%%%%%%%%%%%%%%%%%%%%%%%%%%%%%%%%%%%%%%%%%%%%%%%%%%%%

\clearpage

\xxx {Mode of operation and validation\secl{mode}}

For demonstration purposes we would apply this procedure to a few cases
selected for study.
For the historical cases we can examine \J{d} for times and locations
surrounding the actual tornadoes that did occur.
For validation we will then subjectively compare the maps of \J{d} to
the tornado reports.
We anticipate that high values of \J{d} will be correlated with the
occurrence and intensities of observed tornadoes.
Such a finding will be necessary to consider the method useful.
While forecasts of severe storms are expected to be improved with the
most sophisticated physical parameterizations and the highest possible
resolution, for forecasting precursor situations predominantly forced
by synoptic scale advection, the forecast model requirements are much
less stringent.
A typical \dvar4 setup would use the Weather Research and Forecast
(WRF) model configured with the YSU (Yonsei University) PBL
parameterization \citep{HonND06} and the Anthes Kuo convection scheme,
a single-moment microphysics scheme with multiple liquid and ice water
species, and the RRTM radiation package.
The initial and boundary conditions for \dvar4 case studies are
available eight-times daily at \km[-]{32} resolution from the North
American Regional Reanalysis Project \citep{MesDK+06}.

Calculating the worst case for different areas and times will provide
a measure of threat.
Skilled on-duty forecasters now issue outlooks for medium- and
high-risk convective situations one to three days in advance.
Alternatively, the Short Range Ensemble Forecast (SREF) system can be
used to identify times and areas with non-zero probability of values
of STP exceeding a critical threshold \citep{BriWS+08}.
In either case, detailed exigent forecasts would be made 3--12 hours
in advance for the areas and times at risk.
Typically, in \eqr{JD1} the area $A$ will be a county, the time
interval $\Delta t$ will be one hour centered on a desired forecast
time, and \J{} defined in \eqr{Jboc} will be minimized for all areas
and time intervals that are considered at risk or are in the
neighborhood of areas and time intervals at risk.
This will produce a sequence of maps of $J_d$ (at constant \J{b} as
described in \secr{method}) that we anticipate will prove very useful
to the operational forecaster.

The methodology described here (and in \secr{method}) may be quite
computer intensive, yet operations in a ``warn on forecast'' setting
are quite time sensitive.
There are probably many possible approaches to increase efficiency.
First, \dvar4 can find the minimum faster if it starts with a good
estimate.
Having found the minimum for one location and time, that solution may
prove to be useful for starting the search for neighboring locations
and times.
Second, resources should be concentrated on locations and times most
at risk.
For example if we first minimize \J{} for a large region $A$ and long
time interval $\Delta t$, then we should be able to identify, in that
minimizing solution, smaller regions and time intervals most at risk
(\ie, where STP is large) for further analysis.
And this process would then be iterated down to the smallest scales
required, but just at the most critical locations and times.

\xx {Future context and outlook\secl{future}}

In this paper we discussed XF of the precursors of tornadoes in
an attempt to avoid missed forecasts.
We identified dry-line tornadogenesis
precursors as a relevant but doable high impact weather phenomena.
The concept of XF is new and can be extended in many ways.
In the future, with further developments of WFR \dvar4, new classes of
XF experiments will become possible (\secr[~\ref{sec:future}.]{beyond}).
Additional potential applications are briefly described in
\secr[~\ref{sec:future}.]{NLSV}.

\xxx {Beyond precursors\secl{beyond}}

XF may benefit operational forecasters by identifying minor changes to
model initial conditions that would strongly influence the weather in
their forecast regions.
Also, this technique might help to identify model or analysis system
deficiencies when applied to poorly predicted historical cases.
The sensitivity of severe storms or hurricanes to slight changes in
environmental flow, their complex dynamics, their relatively compact
size, and their impact on lives and property make them interesting
candidates for study from the XF perspective.
But many other situations could be usefully examined this way.
Examples include local extreme weather events, such as heavy
precipitation, strong winds and extreme
temperatures, as well as situations in which a non-linear response in
societal or economic costs results from modest changes to seemingly
innocuous values of the meteorological variables, such as a reduction
in surface temperatures to slightly below freezing in a region of
citrus production.
XF is applicable when the consequences of a weather event, be they
financial, related to human safety, or otherwise, would be significant
whether or not the meteorological situation is considered
``significant''.
Ultimately the technique involves evaluating the reasonableness of
calculated changes to the existing analysis obtained by presenting
\dvar4 with the usual model background field, observations, and a cost
function term related to the specific event of consequence.

Despite these promising opportunities, a number of limitations now
combine to constrain the experiments that could be conducted.
The critical limitations are the initial conditions, models,
background cost function, and impact of nonlinearities.
Because of these limitations we have described a potential
demonstration project to forecast precursor situations predominantly
forced by synoptic scale advection.
The outlook for overcoming these limitations is reviewed here.

\textit{Initial conditions:} Spinning up a realistic severe storm
simulation is fraught with difficulty.
The use of high resolution radar and satellite data to properly
initialize a mesoscale model is an area of ongoing research.
The success of the technique in application to actual exigent weather
phenomena as opposed to precursors of such phenomena will require
greatly improved capability to initialize the relevant structures in
NWP models.

\textit{Model errors:} As described above, the effect of model errors
is not accounted for by XF.
Including additional elements in the control vector can allow for the
possibility of model error.
For example, following the idea of \citet{Tre05}, perturbations can be
introduced at intervals within the \dvar4 interval.
Also see \secr[~\ref{sec:future}.\ref{sec:NLSV}.]{parameter} below.

\textit{Background cost function:} Available background error
covariances may not be appropriate for the particular synoptic
situations of interest.
For background error covariances, XF can make use of an ensemble of
forecasts, such as from a short-range ensemble forecast (SREF) system
\citep[\eg,][]{GriM02}.

\textit{Nonlinearities:} Incremental or not, \dvar4 relies on
linearizing the governing equations, and on the adjoint of the tangent
linear model.
This results in some difficulties.
First, having the correctly coded adjoint is not a guarantee of
success.
For example, linear instabilities can sometimes be present in the
tangent linear and adjoint versions while the corresponding nonlinear
parameterization is well behaved \citep{Mah99}.
That is, a formally correct adjoint model can be computationally
unstable.
Furthermore, such an instability may only be apparent in a stressing
situation, \eg, a cloud resolving severe storm simulation or a high
resolution hurricane simulation.
Second, even at short forecast times, linearization errors can be
large \citep{Tre04}.
In principle the brute force approach can be used.
It now appears that sufficient computational power has become
available to apply this approach to research questions in a case study
context using a detailed mesoscale model \citep{MarX04}.

\xxx {Further applications\secl{NLSV}}

\xxxx {Forecast sensitivity studies}

Variants of XF \dvar4 have been used to examine forecast sensitivity.
Singular vectors are commonly used to identify forecast sensitivity
\citep[\eg,][]{GelBP+98}.
This is normally a linear sensitivity in which the moist processes
are often simplified or neglected, although some authors
\citep[\eg,][]{ReyR03} have examined how singular vectors evolve
nonlinearly once they are determined.
The choice of the norm that measures the size of the singular vector
at the end of the evolution period is analogous to our \J{d}.
For example, singular vectors can be targeted to particular regions
\citep[\eg,][]{PurBP01}.
To do so, the amplification  during the forecast of a small initial
perturbation is maximized with respect some norm, such as kinetic
energy within a particular region.
The singular vector pattern in the initial conditions may then be used
to design a targetted observing strategy.

XF may be considered to be a nonlinear form of singular vector
analysis.
XF might provide further refinement in any study or application that
makes use of singular vectors.
For example the XF approach could be used to diagnose how and why a
critical forecast failed.
For this purpose \J{d} would measure the difference between forecast
and key elements of what in fact happened.
The control vector could be restricted to small regions or just a
subset of the prognostic variables to establish what part of the
initial conditions were to blame for the poor forecast.
Recent work along these lines is reported by \citet{RivLT08} and
\citet{MuZW09}.

\xxxx {Parameter sensitivity\secl{parameter}}

The existing MM5 and WRF \dvar4 systems are geared towards data
assimilation, and the control vector used in the optimization process
contains just the variables defining the initial state of the model.
It is also of interest to consider the sensitivity to other parts of
the modeling system considered constant during the data assimilation
process.
For example, tunable parameters of physics packages, or ``fixed''
external data such as the roughness length ($z_0$) or the antecedent
soil moisture, are not known exactly, and could thus also be included
in the control vector and adjusted alongside the initial conditions.
Such approaches have been used in oceanic
\citep[\eg,][]{SmeO91,She95} and atmospheric
\citep[\eg,][]{WanMZ05,Nord07} data analysis and assimilation.
Within XF, this would address the possibility that plausible changes
in model parameters could, at least in part, lead to a model solution
that maximizes tornado genesis potential.
The associated changes to the software would be isolated to two
separate parts of the \dvar4 system: transformations from the control
vector used in the optimization to model state variables and
parameters, and the physical packages affected by the tunable
parameters or external data under consideration.

\xxxx {Teaching tool}

There are many potential uses of the WRF system for teaching
undergraduate and graduate meteorology.
The XF approach allows one to pose ``what-if'' lab exercises.
For example, what is the minimum and maximum QPF to be expected from a
particular synoptic situation given the uncertainty in IC?
Or how much would SST have to be reduced to change a category 4
hurricane into a category 3 hurricane?

%xx {References}

\normalspacing

% \bibliography{ams-abbrev,current,nwp,question}
% \bibliographystyle{ametsoc}

\end{document}